\documentclass[ALICE,manyauthors]{cernphprep}

\usepackage{multirow}
\usepackage[comma,square,numbers,sort&compress]{natbib}
\usepackage{hyperref}
\usepackage{lineno}
\newcommand{\jpsi}{\rm J/$\psi$}

\begin{document}%

%%%%%%%%%%%%%%%  Title page %%%%%%%%%%%%%%%%%%%%%%%%
\begin{titlepage}
\PHyear{2017}
\PHnumber{245}      % required, will be obtained from PH
\PHdate{13 September}  % required, will be obtained from PH
%

%%% Put your own title + short title here:
\title{Search for collectivity with azimuthal J/$\psi$-hadron correlations\\ in high multiplicity p--Pb collisions at $\sqrt{s_{\rm NN}}$ = 5.02 and 8.16 TeV}
\ShortTitle{Azimuthal \jpsi-hadron correlations in p--Pb collisions}   % appears on right page headers

%%% Do not change the next lines
\Collaboration{ALICE Collaboration\thanks{See Appendix~\ref{app:collab} for the list of collaboration members}}
\ShortAuthor{ALICE Collaboration} % appears on left page headers, do not change

\begin{abstract}
We present a measurement of azimuthal correlations between inclusive \jpsi\ and charged
hadrons in p--Pb collisions recorded with the ALICE detector at the CERN LHC. The \jpsi\ are reconstructed at forward (p-going, 2.03 $<$ y $<$ 3.53) and
backward (Pb-going, $-$4.46 $<$ y $<$ $-$2.96) rapidity via their $\mu^+\mu^-$ decay channel, while the charged hadrons are reconstructed at mid-rapidity
($|\eta|$ $<$ 1.8). The correlations are expressed in terms of associated
charged-hadron yields per \jpsi\ trigger. A rapidity gap of at least 1.5 units is required between
the trigger \jpsi\ and the associated charged hadrons. Possible correlations due to collective effects are assessed by subtracting the
associated per-trigger yields in the low-multiplicity collisions from those in the high-multiplicity collisions.
After the subtraction, we observe a strong indication of remaining symmetric structures at $\Delta\varphi$ $\approx$ 0 and $\Delta\varphi$ $\approx$ $\pi$,
similar to those previously found in two-particle correlations
at middle and forward rapidity. The corresponding second-order Fourier coefficient ($v_2$) in the transverse momentum interval between 3 and 6 GeV/$c$
is found to be positive with a significance of about 5$\sigma$.
The obtained results are similar to the \jpsi\ $v_2$ coefficients measured in
Pb--Pb collisions at $\sqrt{s_{\rm NN}}$ = 5.02 TeV, suggesting a common mechanism at the origin of the \jpsi\ $v_2$.
%Possible implications for the theoretical models of the J/$\psi$ production are discussed.
\end{abstract}
\end{titlepage}
\setcounter{page}{2}

\section{Introduction}

The measurement of angular correlations between particles produced in hadron and nucleus collisions is a powerful tool to study the particle production mechanisms.
Usually the two-particle correlation function is expressed in terms of differences in the azimuthal angle ($\Delta\varphi$) and pseudorapidity ($\Delta\eta$) of the emitted particles.
In minimum-bias proton--proton (pp) collisions, the dominant structures in the correlation function are a near-side peak at $(\Delta\varphi,\Delta\eta)$ $\approx$ $(0,0)$ and an away-side
ridge located at $\Delta\varphi$ $\approx$ $\pi$ and elongated in $\Delta\eta$ \cite{Wang:1992db}. The near-side peak originates from jet fragmentation, resonance decays and femtoscopic correlations. The away-side
ridge results from fragmentation of recoil jets.
In collisions of heavy ions, the two-particle correlation function exhibits additional long-range structures elongated in $\Delta\eta$ \cite{Aamodt:2011by}. These structures are
usually interpreted as signatures of collective particle flow
produced during the hydrodynamic evolution of the fireball.
They are analyzed in terms of the Fourier coefficients of the relative angle distributions.
Assuming factorization, these coefficients are then related to the Fourier coefficients
($v_{\rm n}$) of the particle azimuthal distribution relative to the common symmetry plane
of the colliding nuclei's overlap area. 

The discovery of a near-side ridge in high-multiplicity
pp \cite{Khachatryan:2010gv} and p--Pb \cite{CMS:2012qk} collisions has increased the interest in two-particle
angular correlations in small collision systems. These discoveries were followed by the observation
that the near-side ridge in p--Pb collisions is accompanied by an away-side
one \cite{Abelev:2012ola,Aad:2012gla}.
Long-range structures have also been reported in two-particle correlations in d--Au collisions at RHIC \cite{Adare:2013piz,Adamczyk:2015xjc}.
Further studies using multi-particle correlations have proven that the observed long-range
correlations are of a collective origin \cite{Khachatryan:2015waa,Abelev:2014mda,Aaboud:2017acw}. Moreover, the transverse-momentum and particle-mass dependencies of the $v_{\rm n}$ coefficients in p--Pb collisions have been
found to be similar to those measured in A--A collisions, suggesting a common hydrodynamic origin of the observed correlations \cite{ABELEV:2013wsa,Khachatryan:2014jra}. Alternative interpretations, including
Color-Glass Condensate based models \cite{Dusling:2013qoz} and final-state parton-parton scattering \cite{Bzdak:2014dia}, have also been proposed.
Long-range correlations of forward and backward muons with mid-rapidity hadrons have also been found in p--Pb collisions at a centre-of-mass energy per nucleon pair $\sqrt{s_{\rm NN}}$ $=$ 5.02 TeV \cite{Adam:2015bka}. The results show
that these correlations persist across wide rapidity ranges and extend into the high muon transverse-momentum interval, which is dominated by decays of heavy flavours.

In pp collisions, the \jpsi\ resonance is formed mainly from pairs of ${\rm c}$ and $\bar{\rm c}$ quarks produced in hard scattering reactions during the initial stage of the collision.
The theoretical models describing the \jpsi\ production combine calculations of the production of ${\rm c}\bar{\rm c}$ pairs within a perturbative Quantum Chromodynamics approach with the subsequent non-perturbative formation of the ${\rm c}\bar{\rm c}$ bound state \cite{Andronic:2015wma}.
In p--Pb collisions, the production is affected by the modification of parton distribution functions
inside the nucleus \cite{Eskola:2009uj} as well as possible energy loss and inelastic scattering inside nuclear matter \cite{Qiu:1998rz,Arleo:2012rs}.
In A--A collisions, there are two additional
competing phenomena that influence the \jpsi\ production. First is the suppressed production due to the dissociation of the ${\rm c}\bar{\rm c}$ pairs in the quark-gluon plasma \cite{Matsui:1986dk}. Second is the \jpsi\ enhancement via recombination of charm quarks thermalized in the medium \cite{Thews:2000rj,BraunMunzinger:2000px}. The recombination is expected to become prevalent in central collisions at the LHC energies.

Recently, the ALICE Collaboration has published a precise measurement of the second-order Fourier coefficient, $v_2$, of the azimuthal distribution of the \jpsi\ production in Pb--Pb collisions at $\sqrt{s_{\rm NN}}$ $=$ 5.02 TeV \cite{Acharya:2017tgv}.
The results show significant $v_2$ in central and semi-central collisions. The measured \jpsi\ $v_2$ at low and intermediate transverse momentum
can be qualitatively described by a transport model in which the \jpsi\ azimuthal anisotropy is inherited from that of recombined charm quarks \cite{Du:2015wha,Zhao:2012gc}.
However, at higher transverse momentum the data still indicates significant $v_2$ while the transport model predicts significantly smaller values coming mostly from
path-length dependent suppression in the almond-shaped interaction region of the colliding nuclei and from non-prompt \jpsi\ produced from b-hadron decays assuming thermalized b quarks.
Given these results in Pb--Pb collisions, it is of interest to study the \jpsi-hadron azimuthal correlations also in the
smaller p--Pb system. The recombination of charm quarks, if any, should have much smaller
impact, due to the smaller number of initially produced charm quarks with respect to Pb--Pb collisions. The small system size should not lead to a sizeable path-length dependent suppression. Nevertheless, the study of the \jpsi-hadron azimuthal correlations
could allow to determine whenever \jpsi\ production is affected by the medium possibly created in these collisions \cite{Chen:2016dke,Liu:2013via,Ferreiro:2014bia}.

In this Letter, we present results for long-range correlations between forward (p-going, 2.03 $<$ y $<$ 3.53) and backward (Pb-going, $-$4.46 $<$ y $<$ $-$2.96) inclusive \jpsi\ and mid-rapidity charged hadrons in p--Pb collisions at $\sqrt{s_{\rm NN}}$ $=$ 5.02 and 8.16 TeV.
Inclusive \jpsi\ refers to both prompt \jpsi\ (direct and decays from higher mass charmonium states) and non-prompt \jpsi\ (feed down from b-hadron decays).

\section{Experimental setup and data samples}

A detailed description of the ALICE apparatus can be found in Ref.~\cite{Aamodt:2008zz}. Below, we
briefly describe the detector systems essential for the present analysis.

In the following, $\eta$ and $y_{\rm lab}$ will denote the pseudorapidity and rapidity in the
ALICE laboratory system. The muons are reconstructed in the muon spectrometer
covering the range of $-$4 $<$ $\eta$ $<$ $-$2.5.
The spectrometer contains
a front absorber located between 0.9 and 5 m from the nominal interaction point. The absorber is followed
by five tracking stations, each made of two planes of Cathode Pad Chambers. The third station is
placed inside a dipole magnet with 3 Tm field integral. The tracking stations are followed by an iron wall with a thickness of 7.2 interaction lengths
and two trigger stations, each one consisting of two planes of Resistive Plate Chambers. 

The position of the interaction point is obtained using the clusters reconstructed in the Silicon Pixel Detector (SPD) \cite{Aamodt:2010aa,Adam:2015gka}.
The SPD is located in the central barrel of the ALICE apparatus and operated inside a large solenoidal magnet
providing a uniform 0.5 T magnetic field parallel to the beam line.
The SPD consists of two cylindrical layers which cover $|\eta|$ $<$ 2.0 and $|\eta|$ $<$ 1.4 with respect to the nominal
interaction-point, for the inner and outer layer, respectively.
The associated charged hadrons at mid-rapidity are reconstructed via the so called SPD tracklets, short
track segments formed from the clusters in the two layers of the SPD and the primary vertex \cite{Adam:2015gka}. 

The V0 detector \cite{Abbas:2013taa} consists of two rings of 32 scintillator counters each, covering 2.8 $<$ $\eta$ $<$ 5.1 (V0-A) and
$-$3.7 $<$ $\eta$ $<$ $-$1.7 (V0-C), respectively. It is used for triggering and event-multiplicity estimation.

The data samples presented here were collected during the 2013 and 2016 p--Pb LHC runs. The collision
energy was $\sqrt{s_{\rm NN}}$ = 5.02 and 8.16 TeV for the 2013 and 2016 data samples, respectively.
Part of the 5.02 TeV data were collected during the 2016 p--Pb run. Data with both beam
configurations, namely Pb-nucleus momentum (denoted as Pb--p collisions) or proton momentum (denoted as p--Pb collisions)
oriented towards the muon spectrometer, have been analyzed.
The asymmetric beam energies, imposed by the two-in-one LHC magnet design, resulted in collisions whose nucleon-nucleon
centre-of-mass reference system is shifted in rapidity by 0.465 in the direction of the proton beam with respect
to the ALICE laboratory system.
The data were taken with a trigger that required coincidence of minimum-bias (MB) and
dimuon triggers. The MB trigger was provided by the V0 detector requesting a signal in both V0-A and V0-C rings. Its efficiency is found to be about 98\% \cite{Adam:2014qja}.
The dimuon trigger required at least a pair of opposite-sign track segments in the muon trigger system,
each with a transverse momentum ($p_{\rm T}$) above the threshold of the online trigger algorithm. This threshold
was set to provide 50\% efficiency for muon tracks with $p_{\rm T}$ = 0.5 GeV/$c$.

The collected data samples of p--Pb and Pb--p collisions at 5.02 TeV (8.16 TeV) correspond to integrated
luminosities of 8.1 and 5.8 (8.7 and 12.9) nb$^{-1}$, respectively.
The maximum interaction pile-up probability ranged up to 3\% and 8\% during 2013 and 2016 data taking, respectively.

\section{Event, track and dimuon selection}

\label{sec:selection}

The beam-induced background is rejected by requiring that the timing signals from both rings of the V0
detector are compatible with particles coming from collision events.
Events containing multiple collisions (pile-up) are rejected by requiring one single
interaction vertex reconstructed in the SPD and by exploiting the correlation between the number of clusters in the two
layers of the SPD and the number of the reconstructed SPD tracklets.

The longitudinal position of the reconstructed primary vertex ($z_{\rm vtx}$) is required to be
within $\pm$10 cm from the nominal interaction point. The reconstructed SPD tracklets are selected
by applying a $z_{\rm vtx}$-dependent pseudorapidity cut. The cut is adjusted to exclude
the contribution from the edges of the SPD where the detector acceptance is low. For example,
we select tracklets within $-$1.8 $<$ $\eta$ $<$ 0.5, $-$1.3 $<$ $\eta$ $<$ 1.3 and $-$0.5 $<$ $\eta$ $<$ 1.8
for events with $z_{\rm vtx}$ = 10, 0 and $-$10 cm, respectively.
The contribution
from fake and secondary tracklets is reduced by applying a $|\Delta\Phi|$ $<$ 5 mrad cut on the difference between
the azimuthal angles of the clusters in the two layers of the SPD with respect to the primary vertex.
With this cut, the mean $p_{\rm T}$ of the selected charged hadrons is found to be
approximately 0.75 GeV/$c$ \cite{Adam:2015bka}.

The tracks reconstructed in the muon spectrometer are required to emerge at a radial transverse
position between 17.6 and 89.5 cm from
the end of the front absorber in order to avoid regions with higher material budget.
The tracks reconstructed in the tracking chambers are identified as muons by requiring their matching
with corresponding track segments in the trigger chambers.
Background tracks are removed with a selection on the product of
the total track momentum and the distance of closest approach to the primary vertex in the transverse
plane \cite{Abelev:2012pi}.
The selected dimuons are defined as pairs of opposite-sign muon tracks having
$-$4 $<$ $y_{\rm lab}^{\mu\mu}$ $<$ $-$2.5, transverse momentum $p_{\rm T}^{\mu\mu}$ between 0 and 12 GeV/$c$ and
invariant mass $M_{\mu\mu}$ between 1 and 5 GeV/$c^2$.
Only events with at least one dimuon satisfying these selection criteria are considered.

The data samples are split into multiplicity classes based on the total charge deposited
in the two rings (V0-A and V0-C) of the V0 detector (V0M) \cite{Adam:2014qja}. The high-multiplicity
(low-multiplicity) event class is defined as 0--20\% (40--100\%) of the MB trigger event sample.

\section{Analysis}

\label{sec:analysis}

\begin{figure}[!h]
\begin{center}
\includegraphics[width=0.8\textwidth]{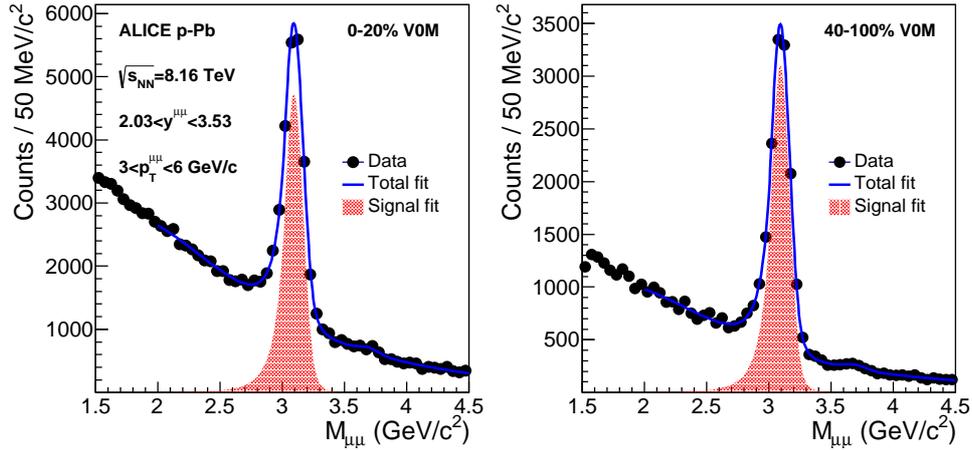}
\caption{\label{fig:fig3} The $M_{\mu\mu}$ distribution in the 3 $<$ $p_{\rm T}^{\mu\mu}$ $<$ 6 GeV/$c$ interval
  fitted with a combination of a CB2 function
  for the signal and a VWG function for the background, for high-multiplicity (left panel) and low-multiplicity
  (right panel) p--Pb collisions at $\sqrt{s_{\rm NN}}$ = 8.16 TeV.}
\end{center}
\end{figure}

The $M_{\mu\mu}$ distribution in each event-multiplicity class and $p_{\rm T}^{\mu\mu}$ bin is fit
with the combination of an extended Crystal Ball (CB2) function for the \jpsi\ signal and a Variable-Width
Gaussian (VWG) function for the background \cite{ALICE-PUBLIC-2015-006}. The tail parameters of the CB2
function were fixed to the values used in \cite{Abelev:2013yxa,ALICE-PUBLIC-2017-001}. The \jpsi\ peak
position and width were obtained
from the fit in the 0--100\% event class and fixed to these values in the other event-multiplicity classes.
Examples of the $M_{\mu\mu}$ fit in the 0--20\% and the 40--100\% event classes in the 3 $<$ $p_{\rm T}^{\mu\mu}$ $<$ 6 GeV/$c$ interval are shown in
Fig.\ref{fig:fig3}.

The angular correlations between \jpsi\ and charged hadrons are
obtained from the associated-particle (SPD tracklets) yields per dimuon trigger.
The yields are defined as

\begin{eqnarray}
  Y^{i}(z_{\rm vtx},M_{\mu\mu},p_{\rm T}^{\mu\mu},\Delta\varphi,\Delta\eta) & = & \frac{1}{N_{\rm trig}^{i}(z_{\rm vtx},M_{\mu\mu},p_{\rm T}^{\mu\mu})}\frac{{\rm d}^2N_{\rm assoc}^{i}(z_{\rm vtx},M_{\mu\mu},p_{\rm T}^{\mu\mu})}{{\rm d}\Delta\varphi{\rm d}\Delta\eta} \nonumber \\
                                                          & = & \frac{1}{N_{\rm trig}^{i}(z_{\rm vtx},M_{\mu\mu},p_{\rm T}^{\mu\mu})}\frac{SE^{i}(z_{\rm vtx},M_{\mu\mu},p_{\rm T}^{\mu\mu},\Delta\varphi,\Delta\eta)}{ME^{i}(z_{\rm vtx},M_{\mu\mu},p_{\rm T}^{\mu\mu},\Delta\varphi,\Delta\eta)},
\label{eq:yield}
\end{eqnarray}

where $N_{\rm trig}^{i}(z_{\rm vtx},M_{\mu\mu},p_{\rm T}^{\mu\mu})$ is the number of dimuons, $N_{\rm assoc}^{i}(z_{\rm vtx},M_{\mu\mu},p_{\rm T}^{\mu\mu})$ is the number of associated SPD tracklets corrected for acceptance and combinatorial effects (as shown in the second line of the equation and described below), $\Delta\varphi$ and
$\Delta\eta$ = $y_{\rm lab}^{\mu\mu} - \eta_{\rm tracklet}$ are the azimuthal angle and (pseudo)rapidity difference
between the trigger dimuon and the associated SPD tracklet. The yields are calculated separately in
each event-multiplicity class (index $i$) and 1 cm-wide $z_{\rm vtx}$ interval. The distribution
\begin{equation}
  SE^{i}(z_{\rm vtx},M_{\mu\mu},p_{\rm T}^{\mu\mu},\Delta\varphi,\Delta\eta)=\frac{{\rm d}^2N_{\rm same}^{i}(z_{\rm vtx},M_{\mu\mu},p_{\rm T}^{\mu\mu})}{{\rm d}\Delta\varphi{\rm d}\Delta\eta} \nonumber
\end{equation}
is the yield of associated SPD tracklets 
from the same event. The distribution
\begin{equation}
  ME^{i}(z_{\rm vtx},M_{\mu\mu},p_{\rm T}^{\mu\mu},\Delta\varphi,\Delta\eta)=\alpha^{i}(z_{\rm vtx},M_{\mu\mu},p_{\rm T}^{\mu\mu})\frac{{\rm d}^2N_{\rm mixed}^{i}(z_{\rm vtx},M_{\mu\mu},p_{\rm T}^{\mu\mu})}{{\rm d}\Delta\varphi{\rm d}\Delta\eta} \nonumber
\end{equation}
is constructed using the event-mixing technique, i.e.\ combining dimuons from one event with
SPD tracklets from other events selected in the same event-multiplicity class and $z_{\rm vtx}$ interval. It serves both to correct for detector acceptance and efficiency and to take into account the combinatorial background. The normalization factor $\alpha^{i}(z_{\rm vtx},M_{\mu\mu},p_{\rm T}^{\mu\mu})$ is
defined as $1/({\rm d}^2N_{\rm mixed}^{i}(z_{\rm vtx},M_{\mu\mu},p_{\rm T}^{\mu\mu})/{\rm d}\Delta\varphi{\rm d}\Delta\eta)$ in the $\Delta\eta$ region corresponding to the maximal
acceptance \cite{Adam:2015bka}.

Within each event-multiplicity class and bin of $M_{\mu\mu}$, $p_{\rm T}^{\mu\mu}$, $\Delta\varphi$ and $\Delta\eta$, the yields $Y^i$ averaged over
$z_{\rm vtx}$ are obtained by fitting the distribution $Y^{i}N_{\rm trig}(z_{\rm vtx})^{i}ME^{i}(z_{\rm vtx})$ to the distribution $SE^{i}(z_{\rm vtx})$. A Poisson likelihood fit is used in order to properly
deal with the cases of low number of tracklets.
Then, the average yields are projected
on the $\Delta\varphi$ axis in the range of
1.5 $<$ $|\Delta\eta|$ $<$ 5 using the method described in \cite{Adam:2015bka}.

\begin{figure}[!h]
\begin{center}
  \includegraphics[width=0.8\textwidth]{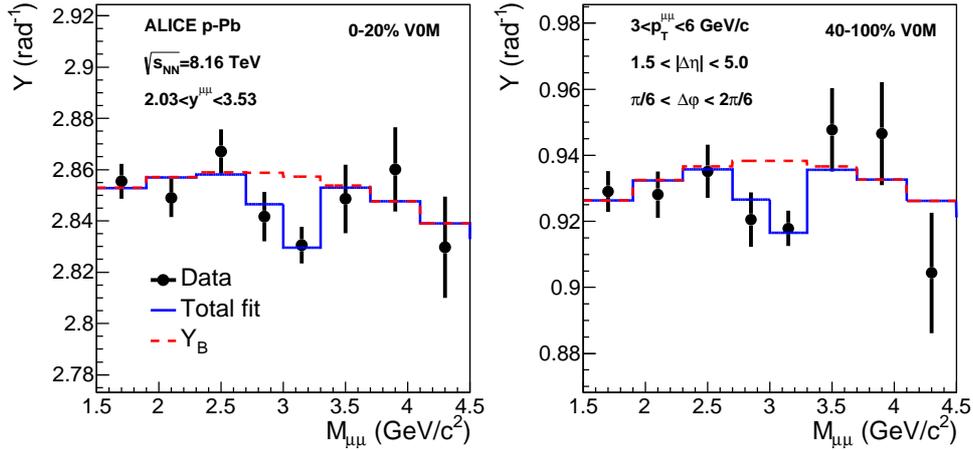}
  \caption{\label{fig:fig6} Example of associated tracklet yields per dimuon trigger in the 3 $<$ $p_{\rm T}^{\mu\mu}$ $<$ 6 GeV/$c$ interval
    for high-multiplicity (left panel) and low-multiplicity (right panel) p--Pb collisions at $\sqrt{s_{\rm NN}}$ = 8.16 TeV.
    The result of the fit with the function from Eq.~(\ref{eq:ext_meth_1}) is represented with the blue solid line.
    The dashed red line corresponds to the associated tracklet yields per background dimuon.
  }
\end{center}
\end{figure}

In order to extract the yields per \jpsi\ trigger, the yields per dimuon trigger in each event-multiplicity
class, $p_{\rm T}^{\mu\mu}$ and $\Delta\varphi$ bins are fit as a function of $M_{\mu\mu}$ using the following superposition
\begin{equation}
Y^{i}(M_{\mu\mu}) = \frac{S}{S+B}Y^{i}_{\rm J/\psi} + \frac{B}{S+B}Y^{i}_{B}(M_{\mu\mu}),
\label{eq:ext_meth_1}
\end{equation}
where $S$ and $B$ are the number of \jpsi\ and the background dimuons in each bin of $M_{\mu\mu}$
obtained from the invariant mass fit (using a CB2 function for the \jpsi\ signal and a
VWG function for the background) described above, $Y_{\rm J/\psi}$ is the associated yield
corresponding to the \jpsi\ trigger and $Y_{B}(M_{\mu\mu})$ is a second-order polynomial function
aimed to describe the associated yields corresponding to the background. The fit range is chosen between
1.5 and 4.5 GeV/$c^2$. Examples of fits in high-multiplicity and low-multiplicity event classes
are shown in Fig.~\ref{fig:fig6}.

\begin{figure}[!h]
\begin{center}
\hfill
\includegraphics[width=0.4\textwidth]{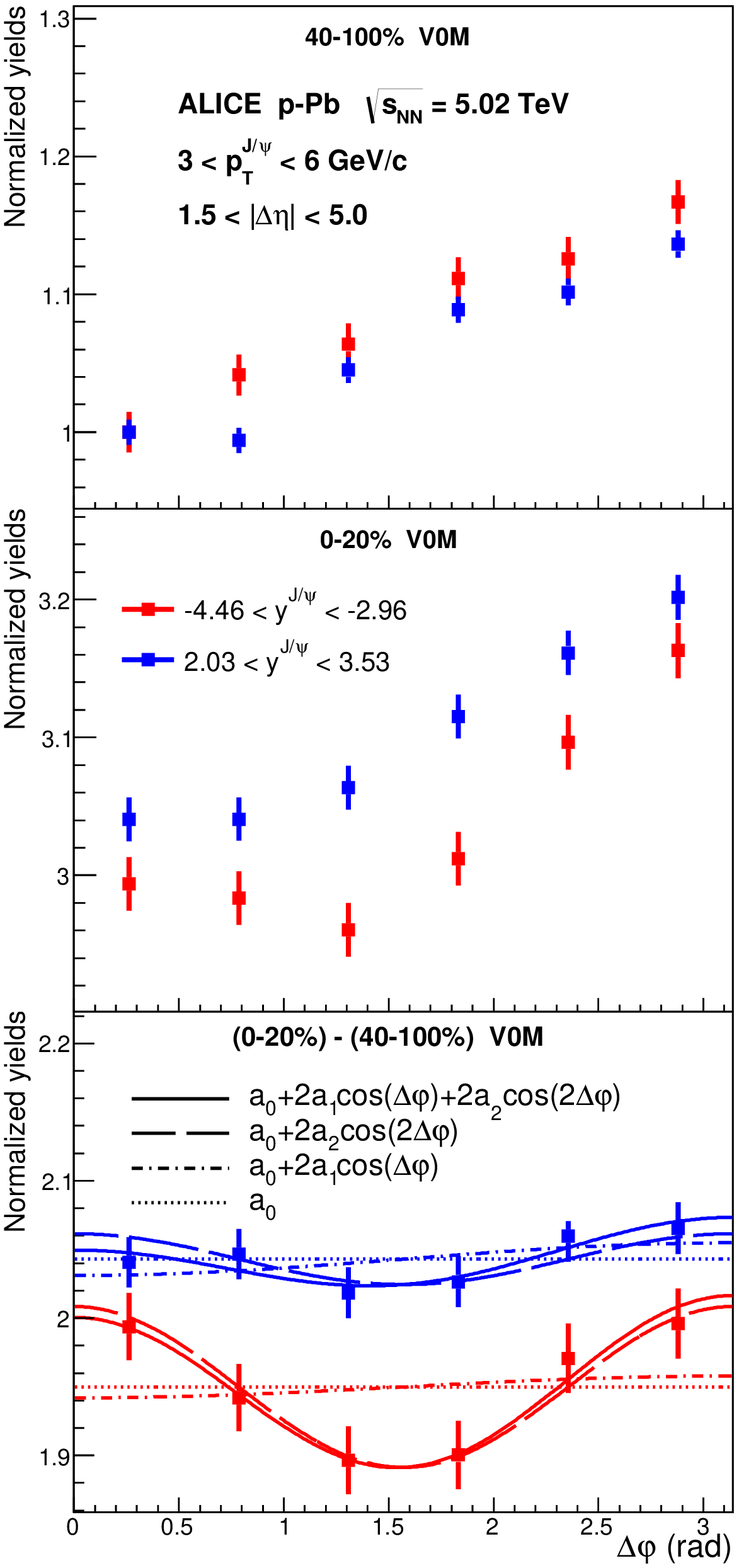}
\hfill
\includegraphics[width=0.4\textwidth]{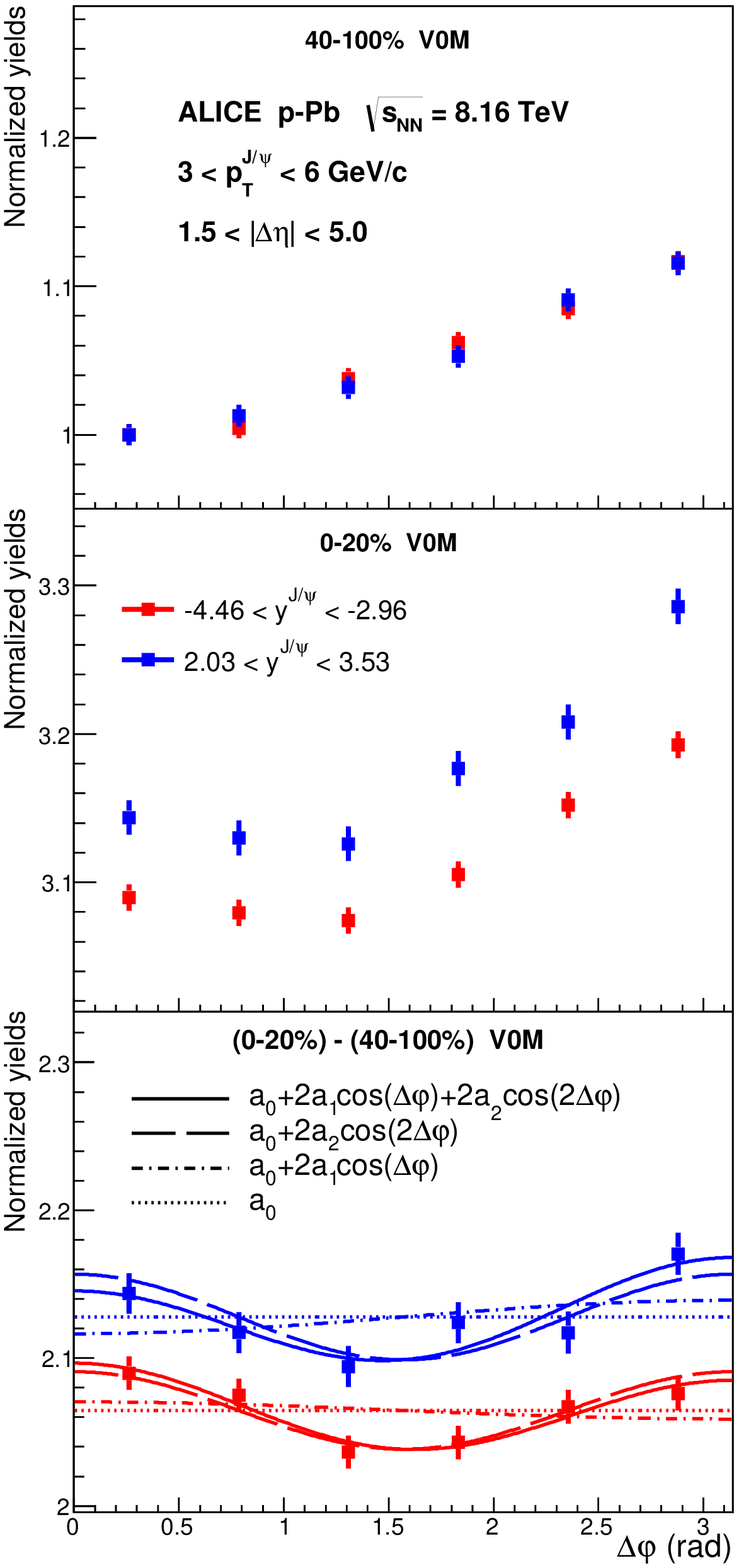}
\hfill
\caption{\label{fig:fig1} Associated tracklet yields per \jpsi\ trigger in 3 $<$ $p_{\rm T}^{\rm J/\psi}$ $<$ $6$
GeV/$c$ in p--Pb and Pb--p collisions at $\sqrt{s_{\rm NN}}$ $=$ 5.02 TeV (left panels) and 8.16 TeV (right panels).
The top and the middle panels correspond to the low-multiplicity and the high-multiplicity event classes, respectively. The bottom
panels show the yields after the subtraction of the low-multiplicity collision yields from the high-multiplicity collision
ones. The solid line represent the fit to the data as described in the text. The dashed, dot-dashed and dotted
lines correspond to the individual terms of the fit function defined in Eq.~(\ref{eq:V2_fit}).
All the yields are normalized to the value in $\Delta\varphi$ $<$ $\pi /6$ in the low-multiplicity (40--100\%) event class.
Only the statistical uncertainties are shown.}
\end{center}
\end{figure}

Figure \ref{fig:fig1} shows the obtained associated tracklet yields per \jpsi\ trigger for p--Pb and Pb--p collisions at $\sqrt{s_{\rm NN}}$ = 5.02 and 8.16 TeV.
As expected, in low-multiplicity collisions we observe a significant correlation
structure on the away side (Fig.~\ref{fig:fig1}, top panels), presumably originating from
the fragmentation of recoil jets.
In high-multiplicity collisions (Fig.~\ref{fig:fig1}, middle panels), a possible enhancement on
both near ($\Delta\varphi$ $\approx$ 0) and away ($\Delta\varphi$ $\approx$ $\pi$) side can be spotted on top of
the away-side structure.
In order to isolate possible correlations due to collective effects between the \jpsi\ and the associated tracklets,
we apply the same subtraction method as in previous
measurements \cite{Abelev:2012ola,Aad:2012gla,ABELEV:2013wsa,Adam:2015bka}, namely
subtracting the $Y_{\rm J/\psi}$ yields in low-multiplicity collisions from those in
high-multiplicity collisions (Fig.~\ref{fig:fig1}, bottom panels).
The subtraction method relies on the assumptions that the jet correlations on the away side
remain unmodified as a function of the event multiplicity and that there are no significant correlations
due to collective effects in low-multiplicity collisions (see discussion in Section \ref{sec:results}).
%This assumption seems to be confirmed by the data, as no sizeable away-side correlation remains after the subtraction.

In order to quantify the remaining correlation structures, the subtracted yields $Y_{\rm J/\psi}^{sub}(\Delta\varphi)$ are fit with
\begin{equation}
a_0 + 2a_1\cos{\Delta\varphi} + 2a_2\cos{2\Delta\varphi}.
\label{eq:V2_fit}
\end{equation}
The second-order Fourier coefficient $V_2\{{\rm J}/\psi-{\rm tracklet,sub}\}$ of the azimuthal correlation
between the \jpsi\ and the associated charged hadrons is finally calculated as $a_2/b_0^{\rm high}$.
The denominator $b_0^{\rm high}=a_0+b_0^{\rm low}$ corresponds to the combinatorial baseline of the high-multiplicity collisions, where the parameter
$b_0^{\rm low}$ is the combinatorial baseline of the low-multiplicity collisions obtained at the minimum of
the per-trigger yields, namely in $\Delta\varphi$ $<$ $\pi /6$. The parameter $b_0^{\rm low}$ is the normalization factor
used in Fig.~\ref{fig:fig1}.
The parameter $a_1$, which describes the strength of the remaining away-side correlation structure, is
found to be compatible with zero in practically all $p_{\rm T}^{\rm J/\psi}$ intervals, in both p--Pb and
Pb--p collisions at both 5.02 and 8.16 TeV.

\begin{figure}[!h]
\begin{center}
  \includegraphics[width=0.5\textwidth]{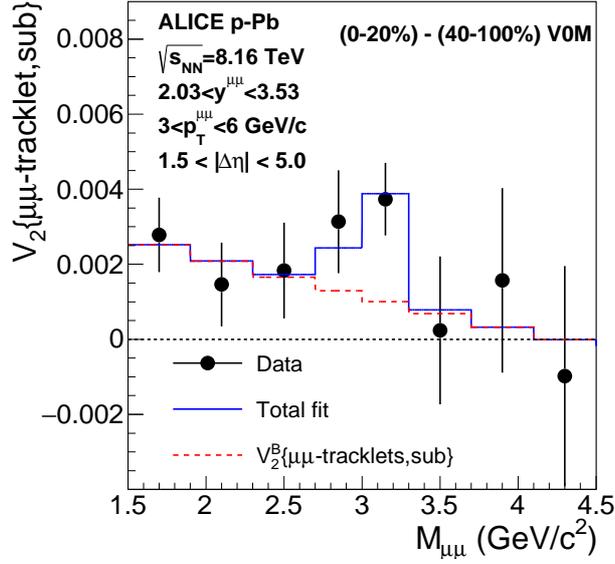}
  \caption{\label{fig:fig5} Example of the fit from
  Eq.~(\ref{eq:ext_meth_2}) in the 3 $<$ $p_{\rm T}^{\mu\mu}$ $<$ 6 GeV/$c$ interval for p--Pb collisions at $\sqrt{s_{\rm NN}}$ = 8.16 TeV. The dashed line corresponds to the $V_2^B\{\mu\mu-{\rm tracklet,sub}\}(M_{\mu\mu})$.}
\end{center}
\end{figure}

As an alternative extraction method, the calculation of $b_0^{\rm low}$, the subtraction of low-multiplicity from
high-multiplicity collision yields and the fit to Eq.~(\ref{eq:V2_fit}) is done in each bin of $M_{\mu\mu}$ separately. Then
the $V_2\{{\rm J}/\psi-{\rm tracklet,sub}\}$ coefficient is extracted by
fitting $V_2\{\mu\mu-{\rm tracklet,sub}\}(M_{\mu\mu})$ with a superposition
similar to the one defined in Eq.~(\ref{eq:ext_meth_1})
\begin{equation}
  V_2\{\mu\mu-{\rm tracklet,sub}\}(M_{\mu\mu})=\frac{S}{S+B}V_2\{{\rm J}/\psi-{\rm tracklet,sub}\} + \frac{B}{S+B}V_2^B\{\mu\mu-{\rm tracklet,sub}\}(M_{\mu\mu}),
  \label{eq:ext_meth_2}
\end{equation}
where the $V_2^B\{\mu\mu-{\rm tracklet,sub}\}(M_{\mu\mu})$ is the second-order Fourier coefficient of the
azimuthal correlation between the background dimuons and associated tracklets.
The background coefficient $V_2^B\{\mu\mu-{\rm tracklet,sub}\}(M_{\mu\mu})$ is parameterized with a second-order
polynomial function. This parameterization is chosen since it reproduces the dimuon $v_2(M_{\mu\mu})$ constructed
from the measured muon $v_2$ coefficient \cite{Adam:2015bka} assuming that
the dominant part of the background is combinatorial.
An example of the $V_2\{\mu\mu-{\rm tracklet,sub}\}(M_{\mu\mu})$ fit is shown in Fig.~\ref{fig:fig5}.

Following the procedure used in Refs.~\cite{Abelev:2012ola,ABELEV:2013wsa,Adam:2015bka}, the $V_2\{{\rm J}/\psi-{\rm tracklet,sub}\}$ coefficient is
factorized into a product of \jpsi\ and charged-hadron $v_2$ coefficients.
Thus, the \jpsi\ second-order Fourier azimuthal coefficient $v_2^{\rm J/\psi}\{{\rm 2,sub}\}$
is obtained as
\begin{equation}
v_2^{\rm J/\psi}\{{\rm 2,sub}\}=V_2\{{\rm J}/\psi-{\rm tracklet,sub}\}/v_2^{\rm tracklet}\{{\rm 2,sub}\},
\label{eq:factorization}
\end{equation}
%where the label '2' inside the curly brackets denotes the coefficient results from two-particle
%correlation analysis approach.
where the $v_2^{\rm tracklet}\{{\rm 2,sub}\}$ is the tracklet
second-order Fourier azimuthal coefficient obtained
by performing the analysis considering SPD tracklets as both trigger and associated particles. The obtained
values of $v_2^{\rm tracklet}\{{\rm 2,sub}\}$ are between 0.067 and 0.069 depending on the beam configuration
and collision energy, with 1--2\% relative statistical uncertainty and 5--6.5\% relative systematic uncertainty.

\section{Systematic uncertainties}

The combined statistical and systematic uncertainties of the measured $v_2^{\rm tracklet}\{{\rm 2,sub}\}$
coefficient for each beam configuration and collision energy are taken as global systematic uncertainties of
the corresponding $v_2^{\rm J/\psi}\{{\rm 2,sub}\}$ coefficients.

All the other systematic uncertainties of the $v_2^{\rm J/\psi}\{{\rm 2,sub}\}$ coefficients are
obtained for each data sample and $p_{\rm T}$ interval separately.
The following sources are considered.

A possible inaccurate correction for the SPD acceptance is assessed by varying
the $z_{\rm vtx}$ range between $\pm$8 and $\pm$12 cm. Systematic uncertainties are assigned only
in the cases of a significant change of the results. The
significance is defined according to the procedure described in Ref.~\cite{Barlow:2002yb}.

The systematic effect related to the uncertainty of the shape of the dimuon background yields $Y_B(M_{\mu\mu})$
is estimated by performing the fit with Eq.~(\ref{eq:ext_meth_1})
using a linear function for the background term and varying the fit range.
The systematic effect coming from the uncertainty of the signal-to-background ratio $S/B$ is
checked by employing various invariant mass fit functions, both for the background and for the \jpsi\ signal.
The maximal difference of the results obtained with the above checks with respect to the default approach
is taken as the corresponding systematic uncertainty.

The uncertainty arising from the employed analysis approach is obtained as the difference between the two extraction
methods described in Section \ref{sec:analysis}.

As described in Section~\ref{sec:analysis}, by default the mixed-event distribution $ME(\Delta\varphi,\Delta\eta)$ is normalized to unity in the $\Delta\eta$
region corresponding to the maximal acceptance. As an alternative approach, normalizing the integral of $ME(\Delta\varphi,\Delta\eta)$
to unity is used. No significant effect on the obtained results is observed and thus no systematic uncertainty is assigned. 

The used event-mixing technique can introduce systematic biases. The event multiplicity distribution of the selected dimuons
(1 $<$ $M_{\mu\mu}$ $<$ 5 GeV/$c^2$) differs from that of the \jpsi\ signal. Since the charged-hadron spectra and the
charged-hadron density as a function of $\eta$ change with event multiplicity \cite{Adam:2014qja}, the non-uniform
(both in the azimuthal and longitudinal directions) SPD acceptance can introduce a bias. The corresponding
systematic uncertainty is evaluated by doing the event mixing in finer event-multiplicity bins.

The non-uniform acceptance of the muon spectrometer coupled to sizeable correlations between the dimuons and
SPD tracklets can bias azimuthally the sample of SPD tracklets used for event mixing. In order to
check for possible effects on our measurement, the event mixing is performed in intervals of azimuthal
angle of the selected dimuons. We observe no
significant systematic effect as the obtained results show negligible deviations with respect to the
results using the default event-mixing technique.

The effect of a possible residual near-side peak is checked by varying the rapidity gap between the
trigger dimuons and associated charged-hadrons from 1.0 to 2.0 units. We observe no indication of
increasing $v_2$ with reduced gap and thus consider the default gap of 1.5 units sufficient to
eliminate any significant residual near-side peak contribution.

As shown in Section \ref{sec:analysis}, the recoil-jet away-side correlation structure in the high-multiplicity
event class is greatly diminished after the subtraction of the low-multiplicity event class.
By default, any remaining away-side structure is supposed to be taken into account by the $\cos{\Delta\varphi}$ term in Eq.~(\ref{eq:V2_fit}).
In order to check for residual effects we proceed in the following way. First, the correlation
function in the low-multiplicity event class is fit with a Gaussian function centered at $\Delta\varphi$ $=$ $\pi$.
Then, the correlation function in the high-multiplicity event class is fit with the function from
Eq.~(\ref{eq:V2_fit}), where the $\cos{\Delta\varphi}$ term is replaced by a Gaussian function with a width
fixed to the value obtained from the fit in the low-multiplicity collisions. No clear signature
of systematic change of the results is seen, except some hints of a possible effect in the highest $p_{\rm T}^{\rm J/\psi}$
interval. Conservatively, we assign systematic uncertainty as the difference with respect
to the default analysis approach.
Since the typical values of
the Gaussian width are around 1 rad, one-sided (negative) systematic uncertainty is assigned.

In Table \ref{tab:syst} we present a summary of the assigned systematic uncertainties of the
$v_2^{\rm J/\psi}\{{\rm 2,sub}\}$ coefficients. No sizeable correlations between
the $p_{\rm T}^{\rm J/\psi}$ intervals are observed and therefore in the following the uncertainties are
considered uncorrelated.
\begin{table}
\begin{center}
\resizebox{\textwidth}{!}{
\begin{tabular}{lcccc}
\hline
\multirow{2}{*}{Source of systematics} & \multicolumn{2}{c}{$\sqrt{s_{\rm NN}}$ $=$ 5.02 TeV} & \multicolumn{2}{c}{$\sqrt{s_{\rm NN}}$ $=$ 8.16 TeV}\\  
                      & p--Pb          & Pb--p          & p--Pb          & Pb--p          \\
\hline
Acceptance correction & 0 to 0.019     & 0 to 0.057     & 0 to 0.011     & 0 to 0.007     \\
Background shape      & 0.007 to 0.013 & 0.015 to 0.056 & 0.011 to 0.013 & 0.003 to 0.012 \\
Extraction method     & 0.003 to 0.015 & 0.010 to 0.040 & 0.002 to 0.011 & 0.008 to 0.018 \\
Event mixing          & 0.003 to 0.015 & 0.004 to 0.025 & 0.002 to 0.008 & 0.004 to 0.012 \\
Residual away-side    & --    & $-$0.030 to 0  & $-$0.018 to 0  & --             \\
jet correlation       & & & & \\
\hline
\multirow{2}{*}{Total}&$+$0.009 to $+$0.024&$+$0.024 to $+$0.084&$+$0.013 to $+$0.019&$+$0.015 to $+$0.021\\
                      &$-$0.009 to $-$0.024&$-$0.024 to $-$0.090&$-$0.015 to $-$0.026&$-$0.015 to $-$0.021\\
\hline
\end{tabular}
}
\end{center}
\caption{Summary of absolute systematic uncertainties of the $v_2^{\rm J/\psi}\{{\rm 2,sub}\}$ coefficients. The uncertainties vary within the indicated ranges
depending on $p_{\rm T}^{\rm J/\psi}$. The values not preceded by a sign represent double-sided uncertainties.}
\label{tab:syst}
\end{table}

Our measurement is for inclusive \jpsi.
The fraction of \jpsi\ from decays of b-hadrons reaches up to about 15\%
at $p_{\rm T}^{\rm J/\psi}$ $\approx$ 6 GeV/$c$ in p--Pb collisions
at $\sqrt{s_{\rm NN}}$ $=$ 5.02 \cite{Aaij:2013zxa} and 8.16 TeV \cite{Aaij:2017cqq}.
Therefore the feed-down contribution is unlikely to influence significantly our results.
In principle, a possible strong multiplicity dependence of the feed-down fraction can potentially affect the
subtraction approach. However, no evidence for such a strong dependence is observed in pp
collisions \cite{Adam:2015ota}.

As additional cross-checks the analysis is done using alternative event-multiplicity estimators, varying
the tracklet $|\Delta\Phi|$ cut, applying a cut on the asymmetry of transverse momentum of
the two muon tracks, removing the pile-up cuts and excluding the SPD regions with non-uniform
acceptance in pseudorapidity. The corresponding results are found to be compatible with those obtained with
the default analysis approach and therefore no further systematic uncertainties are assigned.

\section{Results}
\label{sec:results}

\begin{figure}[!h]
\begin{center}
\includegraphics[width=0.9\textwidth]{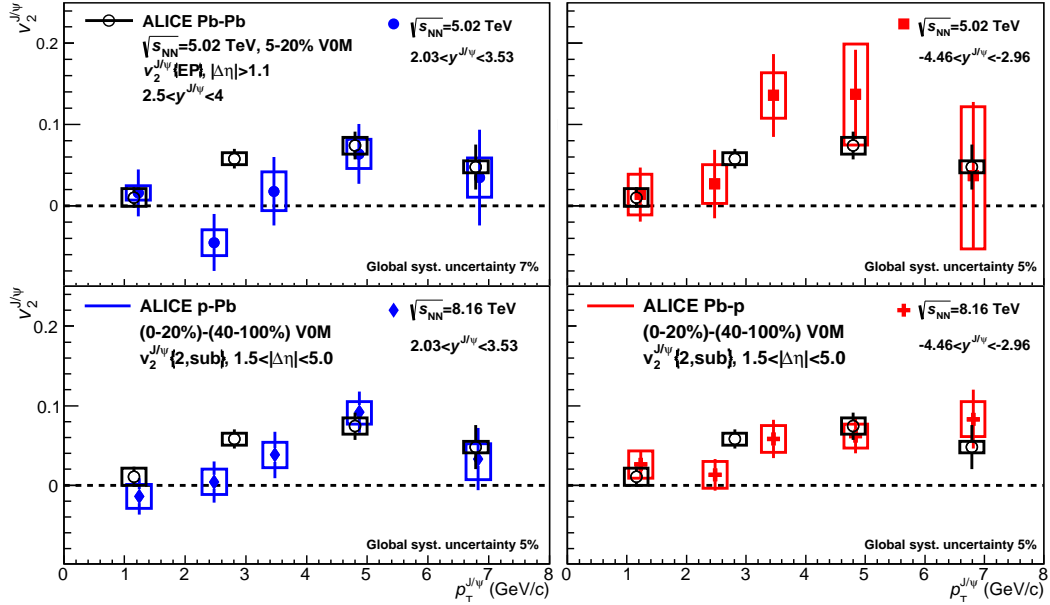}
\caption{\label{fig:fig2} $v_2^{\rm J/\psi}\{{\rm 2,sub}\}$ in bins of $p_{\rm T}^{\rm J/\psi}$ for p--Pb, 2.03 $<$ y $<$ 3.53, (left panels)
and Pb--p, $-$4.46 $<$ y $<$ $-$2.96, (right panels) collisions at $\sqrt{s_{\rm NN}}$ $=$ 5.02 TeV (top panels) and 8.16 TeV (bottom panels).
The results are compared to the
$v_2^{\rm J/\psi}\{{\rm EP}\}$ coefficients measured in central Pb--Pb collisions at $\sqrt{s_{\rm NN}}$ $=$ 5.02
TeV in forward rapidity (2.5 $<$ y $<$ 4) using event plane (EP) based methods \cite{Acharya:2017tgv}. The statistical
and uncorrelated systematic uncertainties are represented by lines and boxes, respectively. The quoted global
systematic uncertainties correspond to the combined statistical and systematic uncertainties
of the measured $v_2^{\rm tracklet}\{{\rm 2,sub}\}$ coefficient.}
\end{center}
\end{figure}

In Fig.~\ref{fig:fig2} we report the measured $v_2^{\rm J/\psi}\{{\rm 2,sub}\}$ coefficients
as a function of $p_{\rm T}^{\rm J/\psi}$ for p--Pb and Pb--p collisions at $\sqrt{s_{\rm NN}}$ = 5.02 and 8.16 TeV.
Up to $p_{\rm T}^{\rm J/\psi}$ of 3 GeV/$c$, no significant deviation from zero is observed for either p--Pb or Pb--p
collisions at the two collision energies. On the contrary, in the $p_{\rm T}^{\rm J/\psi}$ interval between
3 and 6 GeV/$c$, the $v_2^{\rm J/\psi}\{{\rm 2,sub}\}$
is found to be positive although with large uncertainties.
As also shown in Fig.~\ref{fig:fig2}, the $v_2^{\rm J/\psi}$ coefficients in 2.5 $<$ y $<$ 4 in central Pb--Pb collisions
at $\sqrt{s_{\rm NN}}$ $=$ 5.02 TeV reach maximal values in the same $p_{\rm T}^{\rm J/\psi}$ interval \cite{Acharya:2017tgv}.

Two methods are employed in order to obtain the probability that
the $v_2^{\rm J/\psi}\{{\rm 2,sub}\}$ is zero in the 3 $<$ $p_{\rm T}^{\rm J/\psi}$ $<$ 6 GeV/$c$ interval.
In the first method, the $v_2^{\rm J/\psi}\{{\rm 2,sub}\}$ values in the two $p_{\rm T}^{\rm J/\psi}$ intervals
(3 $<$ $p_{\rm T}^{\rm J/\psi}$ $<$ 4 GeV/$c$ and 4 $<$ $p_{\rm T}^{\rm J/\psi}$ $<$ 6 GeV/$c$)
are combined into a weighted average
for each rapidity and collision energy. The obtained probabilities are 0.13\% and 0.13\% (7.8\% and 0.23\%)
for p--Pb and Pb--p collisions, respectively, at $\sqrt{s_{\rm NN}}$ $=$ 8.16 TeV (5.02 TeV). Combining
all eight $v_2^{\rm J/\psi}\{{\rm 2,sub}\}$ values yields a total probability of $1.7\times10^{-7}$. This
corresponds to a 5.1$\sigma$ significance of the measured positive $v_2^{\rm J/\psi}\{{\rm 2,sub}\}$ coefficient.
The second method is Fisher's combined probability test \cite{Fisher1992}. With this method one
obtains probabilities of 0.14\% and 0.23\% (10.3\% and 0.41\%) for p--Pb and Pb--p collisions
at $\sqrt{s_{\rm NN}}$ $=$ 8.16 TeV (5.02 TeV), respectively. The total probability is $1.4\times10^{-6}$ which
corresponds to a 4.7$\sigma$ significance.
In the calculation of the above probabilities, both statistical
and systematic uncertainties of the measured values are taken into account. The global systematic
uncertainty is not taken into account as it is irrelevant in the case of the zero hypothesis.

The analysis method presented in this Letter relies on the assumption that there are no significant
correlations due to collective effects in the low-multiplicity event class.
In case of a presence of such correlations, the measured $V_2\{{\rm J}/\psi-{\rm tracklet,sub}\}$ is equal to
\begin{equation}
  V_2\{{\rm J}/\psi-{\rm tracklet,high}\}-\frac{b_0^{\rm low}}{b_0^{\rm high}}V_2\{{\rm J}/\psi-{\rm tracklet,low}\},
\end{equation}
where $V_2\{{\rm J}/\psi-{\rm tracklet,high}\}$ and $V_2\{{\rm J}/\psi-{\rm tracklet,low}\}$ are the second-order Fourier coefficients of the azimuthal correlation between the \jpsi\ and the associated
charged hadrons in the high-multiplicity and the low-multiplicity collisions, respectively, and
$b_0^{\rm low}/b_0^{\rm high}\approx$1/3 is the ratio of the combinatorial baseline in the low-multiplicity and
high-multiplicity collisions (see Fig.~\ref{fig:fig1}).
As is demonstrated in Ref.~\cite{Aaboud:2016yar}, the
assumption of no significant collective correlations in the low-multiplicity collisions is certainly questionable for light-flavour hadrons. Our data indicates the same, as we observe
a statistically significant increase of the measured values of $v_2^{\rm tracklet}\{{\rm 2,sub}\}$ when subtracting
a lower event-multiplicity, e.g. 60--100\%, class. Ultimately, the value of the $v_2^{\rm tracklet}$ coefficient is found to be about 17\% higher
in case no subtraction is applied. Therefore, replacing the subtracted $v_2^{\rm tracklet}\{{\rm 2,sub}\}$ coefficient in Eq.~(\ref{eq:factorization})
by the non-subtracted coefficient would mean that the $v_2^{\rm J/\psi}$ coefficients are up to 17\% lower
with respect to the measured $v_2^{\rm J/\psi}\{{\rm 2,sub}\}$ coefficients. However, assuming that the
$v_2^{\rm J/\psi}$ coefficients follow the same trend as a function of event multiplicity as the
$v_2^{\rm tracklet}$ coefficient, they would be up to 17\% higher with respect to the measured
$v_2^{\rm J/\psi}\{{\rm 2,sub}\}$ coefficients. Subtracting lower event-multiplicity classes in the
measurement of the $v_2^{\rm J/\psi}\{{\rm 2,sub}\}$ coefficient does not improve the
precision of our measurement, because of the limited amount of \jpsi\ signal in the low-multiplicity
collisions.

\begin{figure}[!h]
\begin{center}
\includegraphics[width=0.5\textwidth]{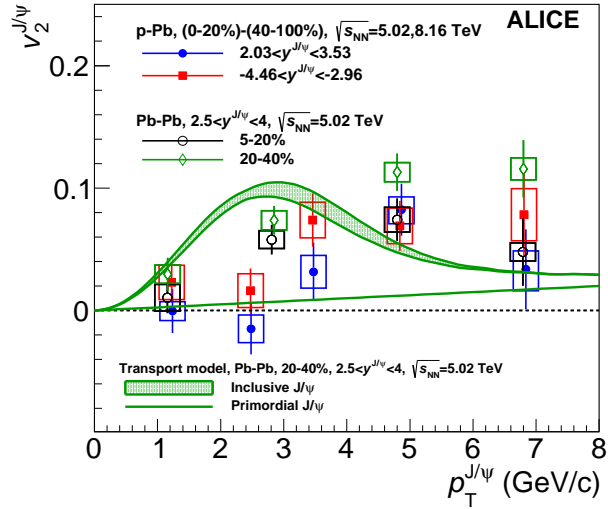}
\caption{\label{fig:fig4} Combined $v_2^{\rm J/\psi}\{{\rm 2,sub}\}$ coefficients in p--Pb and Pb--p collisions
  compared to the results in central and semi-central Pb--Pb collisions at $\sqrt{s_{\rm NN}}$ $=$ 5.02
  TeV \cite{Acharya:2017tgv} and the transport model calculations for semi-central Pb--Pb collisions
  at $\sqrt{s_{\rm NN}}$ $=$ 5.02 TeV \cite{Du:2015wha}. The solid line corresponds to the contribution
  from path-length dependent suppression inside the medium. The band shows the resulting
  $v_2^{\rm J/\psi}$ including also the recombination of thermalized charm quarks and the feed-down
  from b-hadron decays assuming thermalization of b quarks.
}
\end{center}
\end{figure}

The nuclear modification factor of \jpsi\ in p--Pb and Pb--p
collisions \cite{Abelev:2013yxa,ALICE-PUBLIC-2017-001} as well as the charged-particle $v_2$
coefficient \cite{Aad:2015gqa,Khachatryan:2015lva,Khachatryan:2016txc} in pp collisions show
no significant $\sqrt{s_{\rm NN}}$ dependence.  
As seen in Fig.~\ref{fig:fig2}, the measured $v_2^{\rm J/\psi}\{{\rm 2,sub}\}$ coefficients at $\sqrt{s_{\rm NN}}$ = 5.02 and 8.16 TeV also appear to be consistent with each other.
The largest absolute difference between the results at the two collision energies is
observed in Pb--p collisions in the 3 $<$ $p_{\rm T}^{\rm J/\psi}$ $<$ 6 GeV/$c$ interval.
The significance of this difference is rather low (below 1.5$\sigma$), because of
the large uncertainties of the measurement at $\sqrt{s_{\rm NN}}$ = 5.02 TeV.
Hence, the data for the two collision energies are combined as a weighted average taking into account
both statistical and systematic uncertainties.
In Fig.~\ref{fig:fig4}, we present these combined results for p--Pb and Pb--p collisions
together with measurements and model calculations for Pb--Pb collisions at $\sqrt{s_{\rm NN}}$ $=$ 5.02 TeV \cite{Du:2015wha}.

In Pb--Pb collisions, the positive $v_2^{\rm J/\psi}$ coefficients at
$p_{\rm T}^{\rm J/\psi}$ below 3-4 GeV/$c$ are believed to originate from the recombination of
charm quarks thermalized in the medium and are described fairly
well by the transport model \cite{Du:2015wha} (see Fig.~\ref{fig:fig4}). In p--Pb collisions,
the amount of produced charm quarks is small and therefore the contribution from recombination
should be negligible. Our measured values at $p_{\rm T}^{\rm J/\psi}$ $<$ 3 GeV/$c$ are compatible
with zero, in line with this expectation. There is one publication \cite{Liu:2013via} which
suggests that even in p--Pb collisions a sizeable contribution from recombination could
occur due to canonical enhancement effects. The uncertainties of our results
do not allow to confirm or to rule out this scenario.

In Pb--Pb collisions, the measured $v_2^{\rm J/\psi}$ coefficients exceed
substantially the theoretical predictions at $p_{\rm T}^{\rm J/\psi}$ $>$ 4 GeV/$c$, where the main
contribution to $v_2^{\rm J/\psi}$ is expected to come from path-length dependent suppression
inside the medium \cite{Du:2015wha} (see Fig.~\ref{fig:fig4}).
In p--Pb collisions, the medium, if any, has a much smaller size \cite{Adam:2015pya}
and hence very little, if any, path-length dependent effects are expected. In principle, the feed-down
from decays of b-hadrons can give a positive $v_2^{\rm J/\psi}$ at high transverse momentum in case
of a positive b quark $v_2$. However, the latter would have to reach unreasonably high values given
the magnitude of the measured $v_2^{\rm J/\psi}\{{\rm 2,sub}\}$ and the small feed-down fraction.
Despite these considerations, the measured positive $v_2^{\rm J/\psi}$ coefficients would imply
that the \jpsi\ participates in the collective behavior of the p--Pb collision system.
%If interpreted as hydrodynamic flow, the measured positive $v_2^{\rm J/\psi}$ coefficients would imply
%that the \jpsi\ participates in the collective expansion of the matter created in p--Pb collisions.

\section{Summary}

We presented a measurement of the angular correlations between forward and backward \jpsi\ and
mid-rapidity charged hadrons in p--Pb and Pb--p collisions at $\sqrt{s_{\rm NN}}$ $=$ 5.02 and 8.16 TeV. The data
indicate persisting long-range correlation structures
at $\Delta\varphi$ $\approx$ 0 and $\Delta\varphi$ $\approx$ $\pi$, reminiscent of the double ridge
previously found in charged-particle correlations at mid- and forward rapidity.
The corresponding $v_2^{\rm J/\psi}\{{\rm 2,sub}\}$ coefficients in
3 $<$ $p_{\rm T}^{\rm J/\psi}$ $<$ 6 GeV/$c$ are found to be positive
with a total significance of 4.7$\sigma$ to 5.1$\sigma$. The obtained
values, albeit with large uncertainties, are comparable
with those measured in Pb--Pb collisions at $\sqrt{s_{\rm NN}}$ $=$ 5.02 TeV in forward
rapidity.% \cite{Acharya:2017tgv}.
Although the underlying mechanism is not understood, the comparable magnitude of
the $v_2^{\rm J/\psi}$ coefficients at high transverse momentum in p--Pb and
Pb--Pb collisions indicates that this mechanism could be similar in both collision systems.

%%%%% acknowledgements
\newenvironment{acknowledgement}{\relax}{\relax}
\begin{acknowledgement}
\section*{Acknowledgements}
% Version: 2017-09-13

The ALICE Collaboration would like to thank all its engineers and technicians for their invaluable contributions to the construction of the experiment and the CERN accelerator teams for the outstanding performance of the LHC complex.
The ALICE Collaboration gratefully acknowledges the resources and support provided by all Grid centres and the Worldwide LHC Computing Grid (WLCG) collaboration.
The ALICE Collaboration acknowledges the following funding agencies for their support in building and running the ALICE detector:
A. I. Alikhanyan National Science Laboratory (Yerevan Physics Institute) Foundation (ANSL), State Committee of Science and World Federation of Scientists (WFS), Armenia;
Austrian Academy of Sciences and Nationalstiftung f\"{u}r Forschung, Technologie und Entwicklung, Austria;
Ministry of Communications and High Technologies, National Nuclear Research Center, Azerbaijan;
Conselho Nacional de Desenvolvimento Cient\'{\i}fico e Tecnol\'{o}gico (CNPq), Universidade Federal do Rio Grande do Sul (UFRGS), Financiadora de Estudos e Projetos (Finep) and Funda\c{c}\~{a}o de Amparo \`{a} Pesquisa do Estado de S\~{a}o Paulo (FAPESP), Brazil;
Ministry of Science \& Technology of China (MSTC), National Natural Science Foundation of China (NSFC) and Ministry of Education of China (MOEC) , China;
Ministry of Science, Education and Sport and Croatian Science Foundation, Croatia;
Ministry of Education, Youth and Sports of the Czech Republic, Czech Republic;
The Danish Council for Independent Research | Natural Sciences, the Carlsberg Foundation and Danish National Research Foundation (DNRF), Denmark;
Helsinki Institute of Physics (HIP), Finland;
Commissariat \`{a} l'Energie Atomique (CEA) and Institut National de Physique Nucl\'{e}aire et de Physique des Particules (IN2P3) and Centre National de la Recherche Scientifique (CNRS), France;
Bundesministerium f\"{u}r Bildung, Wissenschaft, Forschung und Technologie (BMBF) and GSI Helmholtzzentrum f\"{u}r Schwerionenforschung GmbH, Germany;
General Secretariat for Research and Technology, Ministry of Education, Research and Religions, Greece;
National Research, Development and Innovation Office, Hungary;
Department of Atomic Energy Government of India (DAE), Department of Science and Technology, Government of India (DST), University Grants Commission, Government of India (UGC) and Council of Scientific and Industrial Research (CSIR), India;
Indonesian Institute of Science, Indonesia;
Centro Fermi - Museo Storico della Fisica e Centro Studi e Ricerche Enrico Fermi and Istituto Nazionale di Fisica Nucleare (INFN), Italy;
Institute for Innovative Science and Technology , Nagasaki Institute of Applied Science (IIST), Japan Society for the Promotion of Science (JSPS) KAKENHI and Japanese Ministry of Education, Culture, Sports, Science and Technology (MEXT), Japan;
Consejo Nacional de Ciencia (CONACYT) y Tecnolog\'{i}a, through Fondo de Cooperaci\'{o}n Internacional en Ciencia y Tecnolog\'{i}a (FONCICYT) and Direcci\'{o}n General de Asuntos del Personal Academico (DGAPA), Mexico;
Nederlandse Organisatie voor Wetenschappelijk Onderzoek (NWO), Netherlands;
The Research Council of Norway, Norway;
Commission on Science and Technology for Sustainable Development in the South (COMSATS), Pakistan;
Pontificia Universidad Cat\'{o}lica del Per\'{u}, Peru;
Ministry of Science and Higher Education and National Science Centre, Poland;
Korea Institute of Science and Technology Information and National Research Foundation of Korea (NRF), Republic of Korea;
Ministry of Education and Scientific Research, Institute of Atomic Physics and Romanian National Agency for Science, Technology and Innovation, Romania;
Joint Institute for Nuclear Research (JINR), Ministry of Education and Science of the Russian Federation and National Research Centre Kurchatov Institute, Russia;
Ministry of Education, Science, Research and Sport of the Slovak Republic, Slovakia;
National Research Foundation of South Africa, South Africa;
Centro de Aplicaciones Tecnol\'{o}gicas y Desarrollo Nuclear (CEADEN), Cubaenerg\'{\i}a, Cuba, Ministerio de Ciencia e Innovacion and Centro de Investigaciones Energ\'{e}ticas, Medioambientales y Tecnol\'{o}gicas (CIEMAT), Spain;
Swedish Research Council (VR) and Knut \& Alice Wallenberg Foundation (KAW), Sweden;
European Organization for Nuclear Research, Switzerland;
National Science and Technology Development Agency (NSDTA), Suranaree University of Technology (SUT) and Office of the Higher Education Commission under NRU project of Thailand, Thailand;
Turkish Atomic Energy Agency (TAEK), Turkey;
National Academy of  Sciences of Ukraine, Ukraine;
Science and Technology Facilities Council (STFC), United Kingdom;
National Science Foundation of the United States of America (NSF) and United States Department of Energy, Office of Nuclear Physics (DOE NP), United States of America.    %%%%%%% done by webmaster team
\end{acknowledgement}

%%%%%%%% Bibliography (In case of using bibtex generate the bbl requested by arXiv)
\bibliographystyle{utphys}   % Remember we use title in the biblio
\bibliography{references}
%\input {biblio.tex}  

%%%%%%%%% appendix with author list
\newpage
\appendix
\section{The ALICE Collaboration}
\label{app:collab}
% Collaboration: CERN-LHC-ALICE
% Generation Date is 2017-Sep-13

% How to use:
%%%%%%%%% appendix with author list
%\appendix
%\section{The ALICE Collaboration}
%\label{app:collab}
%\input{Alice_Authorslist_XXXX-Axx-XX.tex}
\begingroup
\small
\begin{flushleft}
S.~Acharya\Irefn{org137}\And 
D.~Adamov\'{a}\Irefn{org94}\And 
J.~Adolfsson\Irefn{org34}\And 
M.M.~Aggarwal\Irefn{org99}\And 
G.~Aglieri Rinella\Irefn{org35}\And 
M.~Agnello\Irefn{org31}\And 
N.~Agrawal\Irefn{org48}\And 
Z.~Ahammed\Irefn{org137}\And 
S.U.~Ahn\Irefn{org79}\And 
S.~Aiola\Irefn{org141}
% ...
\And
\renewcommand\labelenumi{\textsuperscript{\theenumi}~}

\section*{Affiliation notes}
\renewcommand\theenumi{\roman{enumi}}
\begin{Authlist}
\item \Adef{org*}Deceased
\item \Adef{orgI}Dipartimento DET del Politecnico di Torino, Turin, Italy
\item \Adef{orgII}M.V. Lomonosov Moscow State University, D.V. Skobeltsyn Institute of Nuclear, Physics, Moscow, Russia
\item \Adef{orgIII}Department of Applied Physics, Aligarh Muslim University, Aligarh, India
\item \Adef{orgIV}Institute of Theoretical Physics, University of Wroclaw, Poland
\end{Authlist}

\section*{Collaboration Institutes}
\renewcommand\theenumi{\arabic{enumi}~}
\begin{Authlist}
\item \Idef{org1}A.I. Alikhanyan National Science Laboratory (Yerevan Physics Institute) Foundation, Yerevan, Armenia
\item \Idef{org2}Benem\'{e}rita Universidad Aut\'{o}noma de Puebla, Puebla, Mexico
\item \Idef{org3}Bogolyubov Institute for Theoretical Physics, Kiev, Ukraine
\item \Idef{org4}Bose Institute, Department of Physics  and Centre for Astroparticle Physics and Space Science (CAPSS), Kolkata, India
\item \Idef{org5}Budker Institute for Nuclear Physics, Novosibirsk, Russia
\item \Idef{org6}California Polytechnic State University, San Luis Obispo, California, United States
\item \Idef{org7}Central China Normal University, Wuhan, China
\item \Idef{org8}Centre de Calcul de l'IN2P3, Villeurbanne, Lyon, France
\item \Idef{org9}Centro de Aplicaciones Tecnol\'{o}gicas y Desarrollo Nuclear (CEADEN), Havana, Cuba
\item \Idef{org10}Centro de Investigaciones Energ\'{e}ticas Medioambientales y Tecnol\'{o}gicas (CIEMAT), Madrid, Spain
\item \Idef{org11}Centro de Investigaci\'{o}n y de Estudios Avanzados (CINVESTAV), Mexico City and M\'{e}rida, Mexico
\item \Idef{org12}Centro Fermi - Museo Storico della Fisica e Centro Studi e Ricerche ``Enrico Fermi', Rome, Italy
\item \Idef{org13}Chicago State University, Chicago, Illinois, United States
\item \Idef{org14}China Institute of Atomic Energy, Beijing, China
\item \Idef{org15}COMSATS Institute of Information Technology (CIIT), Islamabad, Pakistan
\item \Idef{org16}Departamento de F\'{\i}sica de Part\'{\i}culas and IGFAE, Universidad de Santiago de Compostela, Santiago de Compostela, Spain
\item \Idef{org17}Department of Physics, Aligarh Muslim University, Aligarh, India
\item \Idef{org18}Department of Physics, Ohio State University, Columbus, Ohio, United States
\item \Idef{org19}Department of Physics, Pusan National University, Pusan, Republic of Korea
\item \Idef{org20}Department of Physics, Sejong University, Seoul, Republic of Korea
\item \Idef{org21}Department of Physics, University of Oslo, Oslo, Norway
\item \Idef{org22}Department of Physics and Technology, University of Bergen, Bergen, Norway
\item \Idef{org23}Dipartimento di Fisica dell'Universit\`{a} 'La Sapienza' and Sezione INFN, Rome, Italy
\item \Idef{org24}Dipartimento di Fisica dell'Universit\`{a} and Sezione INFN, Cagliari, Italy
\item \Idef{org25}Dipartimento di Fisica dell'Universit\`{a} and Sezione INFN, Trieste, Italy
\item \Idef{org26}Dipartimento di Fisica dell'Universit\`{a} and Sezione INFN, Turin, Italy
\item \Idef{org27}Dipartimento di Fisica e Astronomia dell'Universit\`{a} and Sezione INFN, Bologna, Italy
\item \Idef{org28}Dipartimento di Fisica e Astronomia dell'Universit\`{a} and Sezione INFN, Catania, Italy
\item \Idef{org29}Dipartimento di Fisica e Astronomia dell'Universit\`{a} and Sezione INFN, Padova, Italy
\item \Idef{org30}Dipartimento di Fisica `E.R.~Caianiello' dell'Universit\`{a} and Gruppo Collegato INFN, Salerno, Italy
\item \Idef{org31}Dipartimento DISAT del Politecnico and Sezione INFN, Turin, Italy
\item \Idef{org32}Dipartimento di Scienze e Innovazione Tecnologica dell'Universit\`{a} del Piemonte Orientale and INFN Sezione di Torino, Alessandria, Italy
\item \Idef{org33}Dipartimento Interateneo di Fisica `M.~Merlin' and Sezione INFN, Bari, Italy
\item \Idef{org34}Division of Experimental High Energy Physics, University of Lund, Lund, Sweden
\item \Idef{org35}European Organization for Nuclear Research (CERN), Geneva, Switzerland
\item \Idef{org36}Excellence Cluster Universe, Technische Universit\"{a}t M\"{u}nchen, Munich, Germany
\item \Idef{org37}Faculty of Engineering, Bergen University College, Bergen, Norway
\item \Idef{org38}Faculty of Mathematics, Physics and Informatics, Comenius University, Bratislava, Slovakia
\item \Idef{org39}Faculty of Nuclear Sciences and Physical Engineering, Czech Technical University in Prague, Prague, Czech Republic
\item \Idef{org40}Faculty of Science, P.J.~\v{S}af\'{a}rik University, Ko\v{s}ice, Slovakia
\item \Idef{org41}Faculty of Technology, Buskerud and Vestfold University College, Tonsberg, Norway
\item \Idef{org42}Frankfurt Institute for Advanced Studies, Johann Wolfgang Goethe-Universit\"{a}t Frankfurt, Frankfurt, Germany
\item \Idef{org43}Gangneung-Wonju National University, Gangneung, Republic of Korea
\item \Idef{org44}Gauhati University, Department of Physics, Guwahati, India
\item \Idef{org45}Helmholtz-Institut f\"{u}r Strahlen- und Kernphysik, Rheinische Friedrich-Wilhelms-Universit\"{a}t Bonn, Bonn, Germany
\item \Idef{org46}Helsinki Institute of Physics (HIP), Helsinki, Finland
\item \Idef{org47}Hiroshima University, Hiroshima, Japan
\item \Idef{org48}Indian Institute of Technology Bombay (IIT), Mumbai, India
\item \Idef{org49}Indian Institute of Technology Indore, Indore, India
\item \Idef{org50}Indonesian Institute of Sciences, Jakarta, Indonesia
\item \Idef{org51}INFN, Laboratori Nazionali di Frascati, Frascati, Italy
\item \Idef{org52}INFN, Sezione di Bari, Bari, Italy
\item \Idef{org53}INFN, Sezione di Bologna, Bologna, Italy
\item \Idef{org54}INFN, Sezione di Cagliari, Cagliari, Italy
\item \Idef{org55}INFN, Sezione di Catania, Catania, Italy
\item \Idef{org56}INFN, Sezione di Padova, Padova, Italy
\item \Idef{org57}INFN, Sezione di Roma, Rome, Italy
\item \Idef{org58}INFN, Sezione di Torino, Turin, Italy
\item \Idef{org59}INFN, Sezione di Trieste, Trieste, Italy
\item \Idef{org60}Inha University, Incheon, Republic of Korea
\item \Idef{org61}Institut de Physique Nucl\'eaire d'Orsay (IPNO), Universit\'e Paris-Sud, CNRS-IN2P3, Orsay, France
\item \Idef{org62}Institute for Nuclear Research, Academy of Sciences, Moscow, Russia
\item \Idef{org63}Institute for Subatomic Physics of Utrecht University, Utrecht, Netherlands
\item \Idef{org64}Institute for Theoretical and Experimental Physics, Moscow, Russia
\item \Idef{org65}Institute of Experimental Physics, Slovak Academy of Sciences, Ko\v{s}ice, Slovakia
\item \Idef{org66}Institute of Physics, Academy of Sciences of the Czech Republic, Prague, Czech Republic
\item \Idef{org67}Institute of Physics, Bhubaneswar, India
\item \Idef{org68}Institute of Space Science (ISS), Bucharest, Romania
\item \Idef{org69}Institut f\"{u}r Informatik, Johann Wolfgang Goethe-Universit\"{a}t Frankfurt, Frankfurt, Germany
\item \Idef{org70}Institut f\"{u}r Kernphysik, Johann Wolfgang Goethe-Universit\"{a}t Frankfurt, Frankfurt, Germany
\item \Idef{org71}Institut f\"{u}r Kernphysik, Westf\"{a}lische Wilhelms-Universit\"{a}t M\"{u}nster, M\"{u}nster, Germany
\item \Idef{org72}Instituto de Ciencias Nucleares, Universidad Nacional Aut\'{o}noma de M\'{e}xico, Mexico City, Mexico
\item \Idef{org73}Instituto de F\'{i}sica, Universidade Federal do Rio Grande do Sul (UFRGS), Porto Alegre, Brazil
\item \Idef{org74}Instituto de F\'{\i}sica, Universidad Nacional Aut\'{o}noma de M\'{e}xico, Mexico City, Mexico
\item \Idef{org75}IRFU, CEA, Universit\'{e} Paris-Saclay, Saclay, France
\item \Idef{org76}iThemba LABS, National Research Foundation, Somerset West, South Africa
\item \Idef{org77}Joint Institute for Nuclear Research (JINR), Dubna, Russia
\item \Idef{org78}Konkuk University, Seoul, Republic of Korea
\item \Idef{org79}Korea Institute of Science and Technology Information, Daejeon, Republic of Korea
\item \Idef{org80}KTO Karatay University, Konya, Turkey
\item \Idef{org81}Laboratoire de Physique Subatomique et de Cosmologie, Universit\'{e} Grenoble-Alpes, CNRS-IN2P3, Grenoble, France
\item \Idef{org82}Lawrence Berkeley National Laboratory, Berkeley, California, United States
\item \Idef{org83}Moscow Engineering Physics Institute, Moscow, Russia
\item \Idef{org84}Nagasaki Institute of Applied Science, Nagasaki, Japan
\item \Idef{org85}National and Kapodistrian University of Athens, Physics Department, Athens, Greece
\item \Idef{org86}National Centre for Nuclear Studies, Warsaw, Poland
\item \Idef{org87}National Institute for Physics and Nuclear Engineering, Bucharest, Romania
\item \Idef{org88}National Institute of Science Education and Research, HBNI, Jatni, India
\item \Idef{org89}National Nuclear Research Center, Baku, Azerbaijan
\item \Idef{org90}National Research Centre Kurchatov Institute, Moscow, Russia
\item \Idef{org91}Niels Bohr Institute, University of Copenhagen, Copenhagen, Denmark
\item \Idef{org92}Nikhef, Nationaal instituut voor subatomaire fysica, Amsterdam, Netherlands
\item \Idef{org93}Nuclear Physics Group, STFC Daresbury Laboratory, Daresbury, United Kingdom
\item \Idef{org94}Nuclear Physics Institute, Academy of Sciences of the Czech Republic, \v{R}e\v{z} u Prahy, Czech Republic
\item \Idef{org95}Oak Ridge National Laboratory, Oak Ridge, Tennessee, United States
\item \Idef{org96}Petersburg Nuclear Physics Institute, Gatchina, Russia
\item \Idef{org97}Physics Department, Creighton University, Omaha, Nebraska, United States
\item \Idef{org98}Physics department, Faculty of science, University of Zagreb, Zagreb, Croatia
\item \Idef{org99}Physics Department, Panjab University, Chandigarh, India
\item \Idef{org100}Physics Department, University of Cape Town, Cape Town, South Africa
\item \Idef{org101}Physics Department, University of Jammu, Jammu, India
\item \Idef{org102}Physics Department, University of Rajasthan, Jaipur, India
\item \Idef{org103}Physikalisches Institut, Eberhard Karls Universit\"{a}t T\"{u}bingen, T\"{u}bingen, Germany
\item \Idef{org104}Physikalisches Institut, Ruprecht-Karls-Universit\"{a}t Heidelberg, Heidelberg, Germany
\item \Idef{org105}Physik Department, Technische Universit\"{a}t M\"{u}nchen, Munich, Germany
\item \Idef{org106}Research Division and ExtreMe Matter Institute EMMI, GSI Helmholtzzentrum f\"ur Schwerionenforschung GmbH, Darmstadt, Germany
\item \Idef{org107}Rudjer Bo\v{s}kovi\'{c} Institute, Zagreb, Croatia
\item \Idef{org108}Russian Federal Nuclear Center (VNIIEF), Sarov, Russia
\item \Idef{org109}Saha Institute of Nuclear Physics, Kolkata, India
\item \Idef{org110}School of Physics and Astronomy, University of Birmingham, Birmingham, United Kingdom
\item \Idef{org111}Secci\'{o}n F\'{\i}sica, Departamento de Ciencias, Pontificia Universidad Cat\'{o}lica del Per\'{u}, Lima, Peru
\item \Idef{org112}SSC IHEP of NRC Kurchatov institute, Protvino, Russia
\item \Idef{org113}Stefan Meyer Institut f\"{u}r Subatomare Physik (SMI), Vienna, Austria
\item \Idef{org114}SUBATECH, IMT Atlantique, Universit\'{e} de Nantes, CNRS-IN2P3, Nantes, France
\item \Idef{org115}Suranaree University of Technology, Nakhon Ratchasima, Thailand
\item \Idef{org116}Technical University of Ko\v{s}ice, Ko\v{s}ice, Slovakia
\item \Idef{org117}Technical University of Split FESB, Split, Croatia
\item \Idef{org118}The Henryk Niewodniczanski Institute of Nuclear Physics, Polish Academy of Sciences, Cracow, Poland
\item \Idef{org119}The University of Texas at Austin, Physics Department, Austin, Texas, United States
\item \Idef{org120}Universidad Aut\'{o}noma de Sinaloa, Culiac\'{a}n, Mexico
\item \Idef{org121}Universidade de S\~{a}o Paulo (USP), S\~{a}o Paulo, Brazil
\item \Idef{org122}Universidade Estadual de Campinas (UNICAMP), Campinas, Brazil
\item \Idef{org123}Universidade Federal do ABC, Santo Andre, Brazil
\item \Idef{org124}University of Houston, Houston, Texas, United States
\item \Idef{org125}University of Jyv\"{a}skyl\"{a}, Jyv\"{a}skyl\"{a}, Finland
\item \Idef{org126}University of Liverpool, Liverpool, United Kingdom
\item \Idef{org127}University of Tennessee, Knoxville, Tennessee, United States
\item \Idef{org128}University of the Witwatersrand, Johannesburg, South Africa
\item \Idef{org129}University of Tokyo, Tokyo, Japan
\item \Idef{org130}University of Tsukuba, Tsukuba, Japan
\item \Idef{org131}Universit\'{e} Clermont Auvergne, CNRS/IN2P3, LPC, Clermont-Ferrand, France
\item \Idef{org132}Universit\'{e} de Lyon, Universit\'{e} Lyon 1, CNRS/IN2P3, IPN-Lyon, Villeurbanne, Lyon, France
\item \Idef{org133}Universit\'{e} de Strasbourg, CNRS, IPHC UMR 7178, F-67000 Strasbourg, France, Strasbourg, France
\item \Idef{org134}Universit\`{a} degli Studi di Pavia, Pavia, Italy
\item \Idef{org135}Universit\`{a} di Brescia, Brescia, Italy
\item \Idef{org136}V.~Fock Institute for Physics, St. Petersburg State University, St. Petersburg, Russia
\item \Idef{org137}Variable Energy Cyclotron Centre, Kolkata, India
\item \Idef{org138}Warsaw University of Technology, Warsaw, Poland
\item \Idef{org139}Wayne State University, Detroit, Michigan, United States
\item \Idef{org140}Wigner Research Centre for Physics, Hungarian Academy of Sciences, Budapest, Hungary
\item \Idef{org141}Yale University, New Haven, Connecticut, United States
\item \Idef{org142}Yonsei University, Seoul, Republic of Korea
\item \Idef{org143}Zentrum f\"{u}r Technologietransfer und Telekommunikation (ZTT), Fachhochschule Worms, Worms, Germany
\end{Authlist}
\endgroup
  %%%%%%% done by webmaster team
\end{document}